\documentclass[aps,prl,twocolumn, superscriptaddress, amsmath]{revtex4-1}
\usepackage{natbib}
\usepackage{graphicx}
\usepackage{bm}
\usepackage{color,soul}
\usepackage{hyperref}
\usepackage{float}
\usepackage{csquotes}
\usepackage{color,soul}
\usepackage{hyperref}
\hypersetup{colorlinks=true, urlcolor=blue, citecolor=blue}
\usepackage{longtable}
\usepackage{tabularx}
\usepackage{adjustbox}
\usepackage{braket}

\setcitestyle{journalcolor= blue}

\usepackage{babel}

\begin{document}
\title{Anisotropic in-plane lattice thermal conductivity in bilayer ReS$_{2}$}
\author{Ashutosh Srivastava}
\altaffiliation{These authors contributed equally to this work.}
\author{Nikhilesh Maiity}
\altaffiliation{These authors contributed equally to this work.}
\author{Abhishek Kumar Singh}
\email {abhishek@iisc.ac.in}
\affiliation{Materials Research Centre, Indian Institute of Science, Bangalore 560012, India}

\date{\today}
\begin{abstract}
The significantly weak interlayer coupling strength and puckered structure provide the novel layer-tolerant and anisotropic features in two-dimensional (2D) ReS$_2$. These unique features offer an opportunity to modulate the optoelectronic, vibrational, and transport properties along different lattice directions in ReS$_2$. Here, using first-principles density functional theory (DFT), we investigated the thermal transport properties of ReS$_2$ in AA and AB stacking orders. The anisotopic ratios for lattice thermal conductivities ($\kappa$) are found to be 1.08 and 1.12 for AA and AB stacking, respectively. This anisotropic nature remains intact even at higher temperatures up to 1000K, demonstrating anisotropic robustness. Lower symmetry in AB stacking leads to higher phonon scattering, which results in lower group velocity, smaller phonon lifetime, and thereby lower $\kappa$ along both directions as compared to AA stacking. The strong breathing and shear Raman modes in AB stacking indicate stronger layer coupling, further confirming the dominant contribution of acoustic modes towards thermal transport. The findings underscore that the stacking-order-driven preferential heat flow in ReS$_2$ and opens up a new dimension for optimizing device performance. 
      
\end{abstract}
   
\maketitle

\section{Introduction}
Two-dimensional (2D) materials have garnered significant attention in optoelectronics that can be tuned with stacking order, layer thickness, and strain, etc.\cite{singh2022origin,castellanos2014isolation,tran2014layer,maity2021anisotropic}
Moreover, compared to the bulk limit, reduced dimensionality and quantum confinement make 2D materials more compelling in the field of thermal transport by affecting phonon scattering mechanisms.\cite{gu2016phonon,gu2018colloquium,song2018two} 
Phononic heat transport properties have been extensively studied in graphene,\cite{nika2009lattice,kong2009first} transition metal dichalcogenides (TMDs), hexagonal boron nitrides (hBN),\cite{lindsay2011enhanced}, and black phosphorus.\cite{luo2015anisotropic} 
Notably, directional heat transport has become a new requirement for modern heat management devices, enhancing their efficiency. In general, these reduced dimensional materials show low out-of-the-plane thermal conductivity compared to in-plane directions.\cite{minnich2015phonon,slack1962anisotropic}
Graphene shows significant in-plane anisotropic phononic heat transport along zigzag (ZZ) and armchair (AC) directions. However, this characteristic starts to disappears as we reach close to bulk limit.\cite{aksamija2011lattice} Additionally, contact resistance, scalability, stability, variability, mechanical strength, and bandgap tunability hinder its practical realization and commercial applicability. Therefore, there is strong urge of a material which could possibly fulfills these requirements.


Recently, ReS$_2$ has gained significant attention within the TMDs family of 2D materials, owing to its unique anisotropic optical and electrical properties.\cite{ho1998absorption,zhou2020stacking} Unlike the other phase of TMDs structures, ReS$_2$ possesses a distorted 1T (1T$'$) triclinic crystal structure. This is because of the presence of additional d valence electrons in Re atoms, leading to the formation of extra Re-Re bonds in ReS$_2$ parallel to the b-axis.\cite{C6NR01569G} For its in-plane anisotropic nature, ReS$_2$ became a distinguished member of the TMD family and is used as a promising polarization-sensitive material in optoelectronic devices. It provides the polarization-dependent excitonic states,\cite{jadczak2019exciton} nonlinear absorption,\cite{meng2018anisotropic} electron transport, and second harmonic generation emission\cite{dhakal2018probing,song2018extraordinary} etc. Owing to the puckered structure of ReS$_2$, it shows in-plane anisotropic properties similar to black phosphorus. However, ReS$_2$ possesses the advantage of being more stable at ambient temperature than black phosphorus, which makes it more suitable for heat management applications. Furthermore, two distinct stacking orders in ReS$_2$ are found to show drastically different features in optical and vibrational properties as well.\cite{zhou2021nonlinear,zhou2020stacking} 

\begin{figure*}[!htp]
    \centering
    \includegraphics[width=1.0\linewidth]{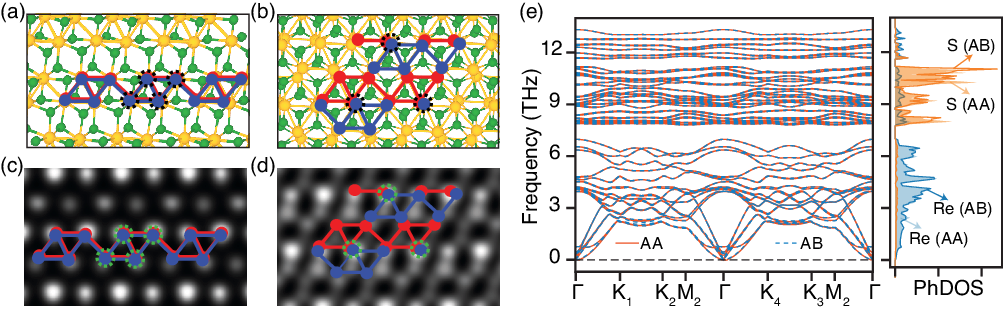}
    \caption{Structure of ReS$_{2}$ in (a) AA stacking and (b) AB stacking. Yellow and green circles represent Re and S atoms, respectively. The dashed line shows the unit cell of the respective stacking. (c)-(d) Simulated STM images of the corresponding AA and AB stacking arrangements. The bottom layer is denoted by the color red, while the upper layer is blue. The green dashed circles denote the luminous dots in crystal structures and STM images. These dots result from the nearly overlapping atoms of two layers. (e) Phonon dispersion and corresponding atom-projected phonon density of states of AA and AB stacking order. Red and blue dashed lines represent the phonon frequencies present in AA and AB stacking along the high-symmetry directions.}
    \label{fig1}
\end{figure*}

Along with the anisotropic properties, the significantly weak interlayer coupling strength makes this material even more special for practical realization at large-scale applications in the field of modern optoelectronics.\cite {Tongay2014} The exfoliation of the controlling number of layers of the 2D materials is a great challenge, and that limits the performance of the nano devices.\cite{C8TC04612C, Rathi2015, PhysRevB.93.075440} Due to the weak coupling strength, the bulk and multilayer ReS$_2$ behave like decoupled monolayers. As a result of this decoupling nature, the special low dimensional characteristic can even be achieved in the bulk and multilayer phases of these materials. For example, the band gap and its nature, i.e., direct band gap, remains intact in monolayer and bulk in this vdW material.\cite{Zhouadv} Furthermore, this weak interlayer coupling in ReS$_2$ leads to the thickness-independent transport properties.\cite{yu2015robust} Moreover, phonon dispersions show a decoupling nature with the increase in the number of layers,\cite{Tongay2014} suggesting the forces among the atoms do not vary significantly with thickness. Experimentally, Jang \textit{et. al.}\cite{jang20173d} have reported the anisotropic lattice thermal conductivity ($\kappa$) and found it to be invariant under the change in the thickness. In addition to this, in-plane $\kappa_l$ parallel and perpendicular to Re-atoms chains are found to be anisotropic and preserved from monolayer to bulk-limit.\cite{Tongay2014,jang20173d} Despite significant advancements, the comprehensive and fundamental understanding behind the stacking order driven anisotropic thermal transport properties in ReS$_2$ is largely overlooked. This gap in research leaves a crucial aspect of these materials’ behavior incomplete. Here, we provide a thorough understanding of $\kappa_l$ along in-plane directions of bilayer ReS$_2$ in AA and AB-stacking order by analyzing the structural, harmonic, and anharmonic properties.


\section{Computational Details}
The first-principles density functional theory has been carried out using Vienna \textit{ab-initio} Simulation Package (VASP)~\cite{kresse1996efficient,kresse1996efficiency}.  Unit cell geometry was optimized using a kinetic energy cut-off of 500 eV. Projector augmented wave (PAW)~\cite{blochl1994projector,kresse1999ultrasoft} pseudopotentials were used to consider electron-ion interactions. Structures were relaxed to the limit such that Hellmann-Feynman forces on each atom become less than 0.005 eV/\AA. The cell parameters were completely relaxed using the conjugate gradient algorithm (CGA) and the Predew-Burke-Ernzehof (PBE) functional with the generalized gradient approximation (GGA). The optB86b-vdW exchange functional~\cite{klimevs2011van,klimevs2009chemical} has been employed to consider weak van der Waal interactions.
A vacuum of $\sim$20{\AA} was used along the z-axis to avoid any interactions between periodic images. The scanning tunneling microscopy (STM) images for AA and AB stacking have been simulated from the partial charge densities. Harmonic second-order interatomic force constants (IFCs) were calculated using the finite-displacement method implemented in the Phonopy code~\cite{togo2015first}. Anharmonic IFCs were calculated by considering the terms up to the third order using phono3py~\cite{phono3py}. Harmonic and anharmonic IFCs were computed by considering supercell size of 5$\times$5$\times$1 and 4$\times$4$\times$1, respectively, with an energy convergence criterion of 10$^{-8}$ eV. Non-analytical term corrections to the dynamical matrix have been considered to take care of long-range interactions by calculating Born effective charges~\cite{PhysRevB.55.10355}. The calculated 2$^{nd}$ order (harmonic) and 3$^{rd}$ (anharmonic) were then used to calculate lattice thermal conductivity ($\kappa_{l}$) by solving the phonon Boltzmann transport equation (PBTE)~\cite{phono3py,chaput2013direct}. A converged q-grid of 19$\times$19$\times$1 has been used to compute $\kappa_{l}$.

\section{Results and Discussions}
\subsection{Stacking orders stability analysis and comparison}
ReS$_2$ exhibits a distorted 1T (1T$'$) lattice structure having two symmetrically inequivalent Re and S atoms. The extra one valence \textit{d} electron in other transition metals, unlike Mo or W, leads to the formation of the Re-Re chain along the b-direction. The crystal structure of ReS$_{2}$ for AA and AB stacking are shown in Figures~\ref{fig1}(a) and (b), respectively.
The optimized lattice parameters and interlayer spacing corresponding to AA and AB stacking orders are mentioned in Table~\ref{tab1} and are in agreement with the experimental observations. In-plane lattice parameters deviate slightly from their monolayer counterpart.\cite{zong2018electric}
\begin{table}[!htbp]
\caption{Lattice parameters and interlayer distance (l$_w$) corresponding to AA and AB stacking order in ReS$_2$.} 
\centering 
\begin{tabular}{|c|c|c|c|c|c|} 
\hline
 & a (\AA) & b (\AA) & l$_w$ (\AA) & $\alpha$,$\beta$ ($^\circ$) & $\gamma$ ($^\circ$) \\ [1ex] 
\hline 
AA  &  6.38  &  6.48 & 2.70 & 90 & 118.84 \\ [1ex] \hline
AB  &  6.38  &  6.48 & 2.59 &  90 &  118.84 \\ [1ex] \hline
\end{tabular}
\label{tab1} 
\end{table}
Moreover, the stacking order does not alter in-plane lattice parameters and crystallographic angles (see Table~\ref{tab1}), but the interlayer distance gets altered by $\sim$5\%, demonstrating the difference in interlayer coupling strength between the two stacking orders. For AA stacking, the top layer sits on the bottom layer with a negligible displacement. In AB stacking, the top layer is displaced by 2.5 {\AA} along the a-axis. The simulated images for AA and AB stacking are well matched (see Figures~\ref{fig1}(c) and (d)) with previously reported experimental HR-TEM images.\cite{zhou2020stacking} Although the two stackings have negligible energy differences due to the significantly weak vdW interlayer coupling strength, the finite barrier height of value 0.36 eV makes establishing two distinct stacking orders possible. As discussed, AA and AB stacking have comparable ground-state energies; however, atomic vibrations around the equilibrium positions could lead to a change in energies, which might destabilize the system. Hence, it is essential to analyze the system's dynamics in response to atomic displacement. Phonon dispersion curves of the AA and AB stacking are shown in Figure~\ref{fig1}(e), illustrating that both stacking orders are dynamically stable. The individual phonon dispersion curves corresponding to AA, AB stacking, and the monolayer have been shown in Figure S1(a-c), respectively.
\begin{figure}[!htbp]
    \centering
    \includegraphics[width=1.0\linewidth]{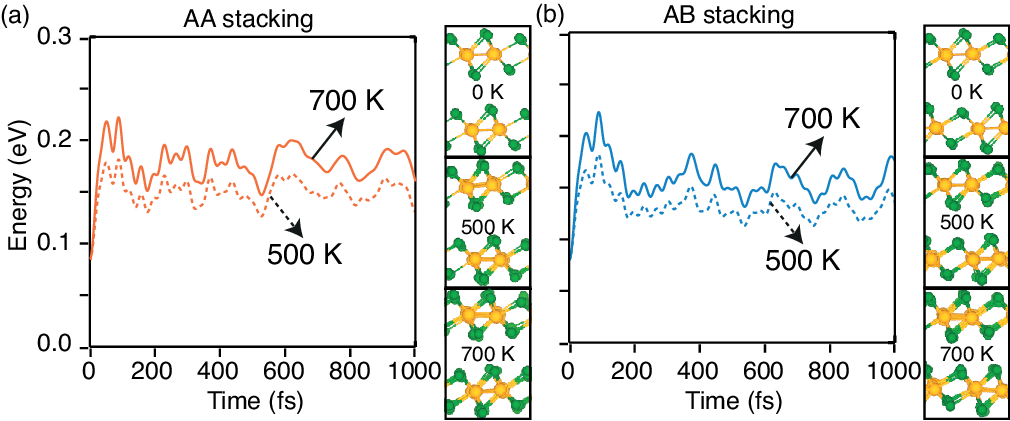}
    \caption{The evolution of energy with time under the NVT ensemble with 500 and 700 K temperatures and corresponding final snapshots (right panel) (a) AA stacking and (b) AB stacking ReS$_2$ using AIMD simulation.}
    \label{fig2}
\end{figure}
There is a minor difference in phonon frequencies observable between the acoustic and low-lying optical modes of the two stacking configurations. Therefore, it is likely that these small differences will not lead to significant deviation in their phononic heat transfer ability. Further, the atom-projected phonon density of states (PhDOS) reveals that the higher phonon frequencies originated from S-atoms, whereas lower frequencies from comparatively heavier Re-atoms. Moreover, the phonon dispersion of these stackings is almost identical to their monolayer counterpart.\cite{zhou2021nonlinear} 

The thermodynamic stability was also tested using AIMD simulation with an NVT ensemble at temperatures 500 and 700 K, as shown in Figures~\ref{fig2}(a) and (b). No significant variation of total energy and forces acting on the atoms is observed in both cases. The energy, electronic, and phonon band structures are very similar in both stacking orders. As their atomic arrangements are completely different in the different stacking orders, low-frequency Raman modes, which are the signature of any stacking order, are totally different.\cite{zhou2020stacking} The Raman spectra show that mode I is located at 136.28 and 131.55 cm$^{-1}$ of AA and AB stacking order, respectively. Furthermore, ultra-low-frequency Raman modes, shearing modes (S$_{parallel}$ and S$_{perpendicular}$), and breathing modes (B) also exhibit distinct behavior, indicating the presence of two distinct stacking orders (see Figure S2). All three modes have higher energy for AB stacking than AA stacking, indicating relatively higher interlayer coupling strength (corresponding lower interlayer distance) in AB stacking. 
These analyses suggest that, despite the differing atomic arrangements in the two stacking orders, they are found to be thermodynamically stable over a wide range of temperatures, thereby maintaining their structural features. 

\subsection{Lattice Thermal Conductivity}
\begin{figure}[!t]
    \centering
    \includegraphics[width=1.0\linewidth]{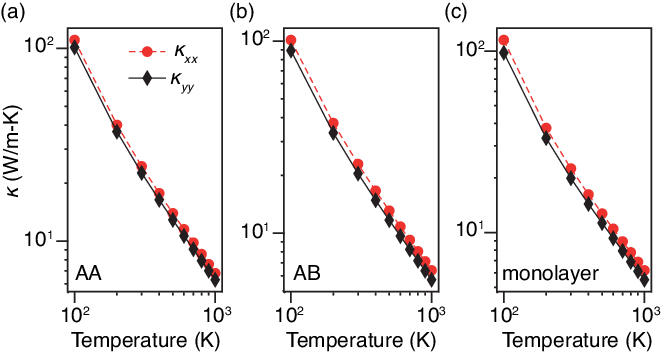}
    \caption{Lattice thermal conductivity of ReS$_{2}$ in (a) AA stacking, (b) AB stacking, and (c) monolayer along the in-plane directions. Circle and diamond represent the lattice thermal conductivity values along and x and y-directions, respectively.}
    \label{fig3}
\end{figure}

Having analyzed the structures representing the two thermodynamically stable stacking orders, we next evaluated the $\kappa$ arising from such structural variations and compared it with the monolayer counterpart. The calculated $\kappa$ values along the two in-plane directions ($\kappa_{xx}$ and $\kappa_{yy}$) corresponding to AA, AB-stacking, and monolayer are shown in Figures~\ref{fig3}(a) to (c), respectively. Evidently, all the structures show anisotropic $\kappa$ having $\kappa_{yy}$ $>$ $\kappa_{xx}$. This variation can be attributed to the inhomogeneous structural distortion along the in-plane conducting directions.
Comparing the $\kappa_{xx}$ and $\kappa_{yy}$ values among AA, AB, and monolayer suggests that AA stacking shows the highest $\kappa$ values and the least by monolayer. The calculated values at 300K are mentioned in Table~\ref{tab2}. Unlike other TMDs, bilayer of ReS$_2$ stacked in two different stable stacking orders shows a marginal variation from its monolayer counterpart. These variations are quantified to be $\sim$8\% (16.8\%) and $\sim$1.60\% (5.59\%) of the AA and AB stacking orders from the monolayer along xx-direction (yy-direction). It indicates that stacking order plays a role in determining $\kappa$ of ReS$_2$. However, the impact is limited to a maximum variation of $\sim$16{\%} on $\kappa$. 
\begin{table}[!htbp]
\caption{Lattice thermal conductivity of AA, AB, and monolayer ReS$_2$ at 300 K.} 
\centering 
\begin{tabular}{|c|c|c|c|} 
\hline
 & $\kappa_{xx}$ (W/m-K) & $\kappa_{yy}$ (W/m-K)  \\ [1ex] 
\hline 
AA  &  24.44  &  22.58   \\ [1ex] \hline
AB  &  22.89  &  20.42   \\ [1ex] \hline
Monolayer  &  22.53  &  19.34   \\ [1ex] \hline
\end{tabular}
\label{tab2} 
\end{table}

\begin{figure}[!htbp]
    \centering
    \includegraphics[width=1.0\linewidth]{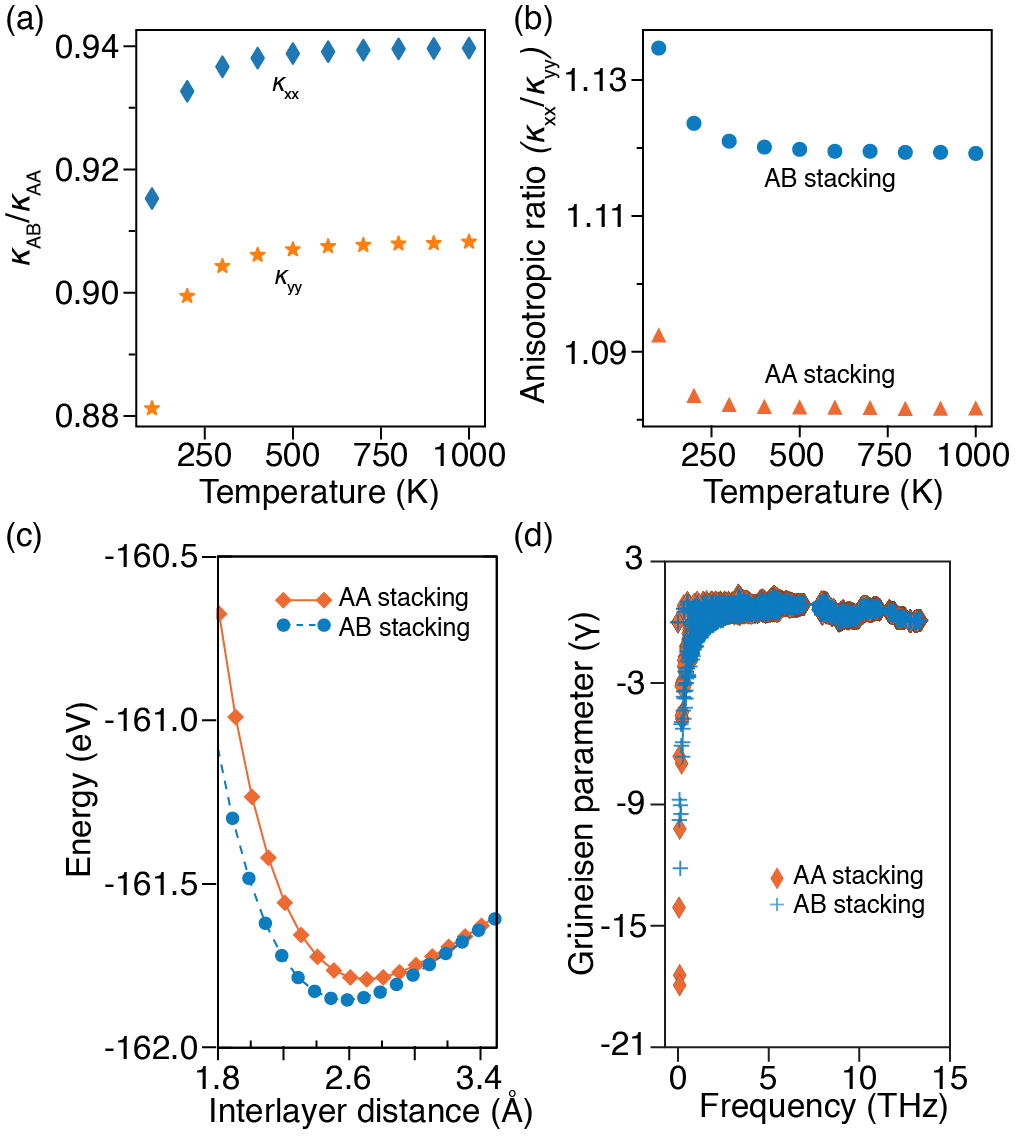}
    \caption{(a) Comparison lattice thermal conductivity between the AB and AA-stacking ($\kappa_{AB}$/$\kappa_{AA}$) orders along two in-plane directions, (b) Anisotropic ratio of lattice thermal conductivity along in-plane directions corresponding to AA and AB-stackings, (c) Energy as a function of interlayer distance of AA and AB-stacking orders, and (d) Gr\"{u}neisen parameters as a function of phonon frequency of AA and AB stacking order.}
    \label{fig4}
\end{figure}

To further describe the characteristics of $\kappa$ along the in-plane directions, anisotropic ratios present within and between AA and AB stacking were calculated.    
The variation ratio of $\kappa_{xx}$ (blue diamonds) and $\kappa_{yy}$ (orange asterisk) between AA and AB stacking up to 1000 K has been shown in Figure~\ref{fig4}(a). The maximum variation between the two stacking orders occurs at temperatures up to 200 K. Thereafter, it gets saturated to $\sim$0.94 and $\sim$0.90 for $\kappa_{xx}$ and $\kappa_{yy}$, respectively. Higher deviation in the $\kappa_{AB}$/$\kappa_{AA}$ ratio along the xx or yy-directions from 1 implies, significant change in $\kappa$ values along that direction. Therefore, this analysis suggests that the yy-direction is more sensitive to the stacking order than the xx-direction in the two stacking orders discussed. Furthermore, to get an idea regarding the anisotropic nature of $\kappa$ within the AA and AB stacking between the in-plane directions, we calculated $\kappa_{xx}$/$\kappa_{yy}$ values as a function of temperatures from 100 to 1000 K, as shown in Figure~\ref{fig4}(b). The values suggest that with an increase in temperature, the anisotropic nature of $\kappa$ along the in-plane directions remains intact, demonstrating temperature-independent anisotropic robustness. AB stacking shows a higher anisotropy ($\sim$1.12) than AA stacking ($\sim$1.08) at temperatures greater than 200 K. This substantiates that the variation in the stacking order influences the phonon-phonon scattering in ReS$_2$. 
Thereby, it indicates that the anisotropies of the phonon-phonon scattering strength are maintained, which preserves the anisotropic nature in the respective stacking orders.   


To understand the interlayer coupling strength between the AA and AB stacking order, we calculated the total energy of the system by varying the interlayer distance as shown in Figure~\ref{fig4}(c). Compared to MoS$_2$,\cite{Tongay2014} both the stacking orders show comparatively very shallower depth in ReS$_2$, although there is a small difference between them. This indicates that the interlayer coupling strength of ReS$_2$ is significantly weaker than that of MoS$_2$, indicating more prominent layer-tolerant behaviour in ReS$_2$.
Moreover, AA stacking exhibits a slightly shallower depth, possessing higher energy than AB stacking. This demonstrates that in AA stacking, the layers are comparatively weakly coupled to each other compared to AB stacking, and this is the reason behind the relatively larger interlayer distance in AA stacking as given in Table~\ref{tab1}. The relatively weaker interlayer coupling strength of AA stacking suggests a more layer-independent nature than that of AB stacking. Hence, the variation in $\kappa$ cannot be solely explained by interlayer coupling strength in ReS$_2$.
Therefore, we further calculated the Gr\"{u}neisen parameter ($\gamma$) that provides insight into the anharmonic characteristics of the structures. The calculated $\gamma$ as a function of phonon frequencies has been plotted in Figure~\ref{fig4}(d) for AA (diamond symbols) and AB (plus symbols) stacking.
The $\gamma$ values are identical (overlapping each other) for both stackings, except for the low-frequency range of 0-1 THz. Thereby suggesting that the ReS$_2$ in AA stacking is more anharmonic, but the anharmonicity is limited to ultra-low phonon frequencies. This partially contradicts the $\kappa$ behavior, where AA stacking shows higher $\kappa$ values than AB stacking, and further analysis of harmonic and anharmonic properties is required to comprehend these characteristics.        

\subsection{Phonon group velocity and lifetime}
In AA stacking, the interlayer coupling strength is weaker, meaning the layers interact less strongly with each other. This weaker interaction leads to larger anharmonicity, which refers to the deviation from perfect harmonic motion in the atomic vibrations. Despite these factors, AA stacking exhibits relatively higher $\kappa$ values compared to AB stacking.    
\begin{figure}[!htbp]
    \centering
    \includegraphics[width=1.0\linewidth]{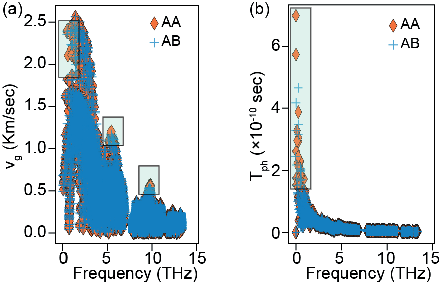}
    \caption{Variation of (a) average phonon group velocity and (b) phonon lifetime as a function of phonon frequencies in the AA and AB stacking.}
    \label{fig5}
\end{figure}
To comprehend this discussion, we further calculated phonon group velocity and phonon lifetime ($\tau_{ph}$) with respect to phonon frequencies. The Phonon group provides the information on the movement of phonons inside the materials, whereas $\tau_{ph}$ estimates the time during which a phonon lasts and carries heat efficiently.\cite{snyder2011complex,mukherjee2022recent} The variation of average phonon group velocity  (v$_g$) and phonon lifetime as a function of phonon frequencies is shown in Figures~\ref{fig5} (a) and (b), respectively. The v$_g$ and $\tau_{ph}$ values corresponding to AA and AB stacking order are indicated by diamond and plus symbols, respectively. Relatively higher $v_g$ and $\tau_{ph}$ values are observed in AA stacking than in AB stacking as visualized by the rectangular boxes in Figures~\ref{fig5}(a) and (b). Thereby, even though AA stacking exhibits weaker interlayer coupling strength, leading to relatively higher anharmonicity compared to AB stacking, it still demonstrates higher $\kappa$ values. This is due to the higher v$_g$ and $\tau_{ph}$ in AA stacking. These factors suggest that phonons move faster and transport heat for a longer period without being scattered, resulting in comparatively higher $\kappa$ values than in AB stacking.

\section{Conclusion}
In conclusion, we have investigated the stacking-driven thermal transport properties of ReS$_2$. 
This study reveals the anisotropic nature of thermal transport resulting from its puckered structure. Analysis of harmonic and anharmonic parameters validates anisotropic characteristics. 
The interlayer coupling strength in ReS$_2$ is extremely weak, leading to variations in $\kappa$ that are nearly independent of the stacking order. Despite this weak interlayer coupling, AA stacking exhibits relatively higher phonon group velocity and phonon lifetime, contributing to a marginally greater $\kappa$ compared to AB stacking. These differences can be attributed to variations in the phonon spectrum, particularly the atomic contributions to the phononic frequencies. Overall, while the stacking order in ReS$_2$ does not drastically alter the thermal conductivity due to the weak interlayer coupling, the higher phonon group velocity and lifetime in AA stacking result in slightly better heat dissipation. 
Additionally, AB stacking shows greater anisotropy compared to AA stacking, which could be advantageous for applications requiring tunable $\kappa$ based on specific needs. 


\section*{Acknowledgements}
The authors thank the Materials Research Centre (MRC), the Solid State and Structural Chemistry Unit (SSCU), and the Indian Institute of Science for their computational facilities. A.S. acknowledges the DST-INSPIRE fellowship [IF190068]. The authors acknowledge the support from the Institute of Eminence (IoE) scheme of the Ministry of Human Resource Development, Government of India.

%

\newpage
\onecolumngrid
\begin{center}
   \textbf{\Large Supplementary Material}
\end{center}

\renewcommand{\thefigure}{S\arabic{figure}}
\setcounter{figure}{0}

\section{Phonon dispersion}
\begin{figure*}[!htbp]
    \centering
    \includegraphics[width=1.0\linewidth]{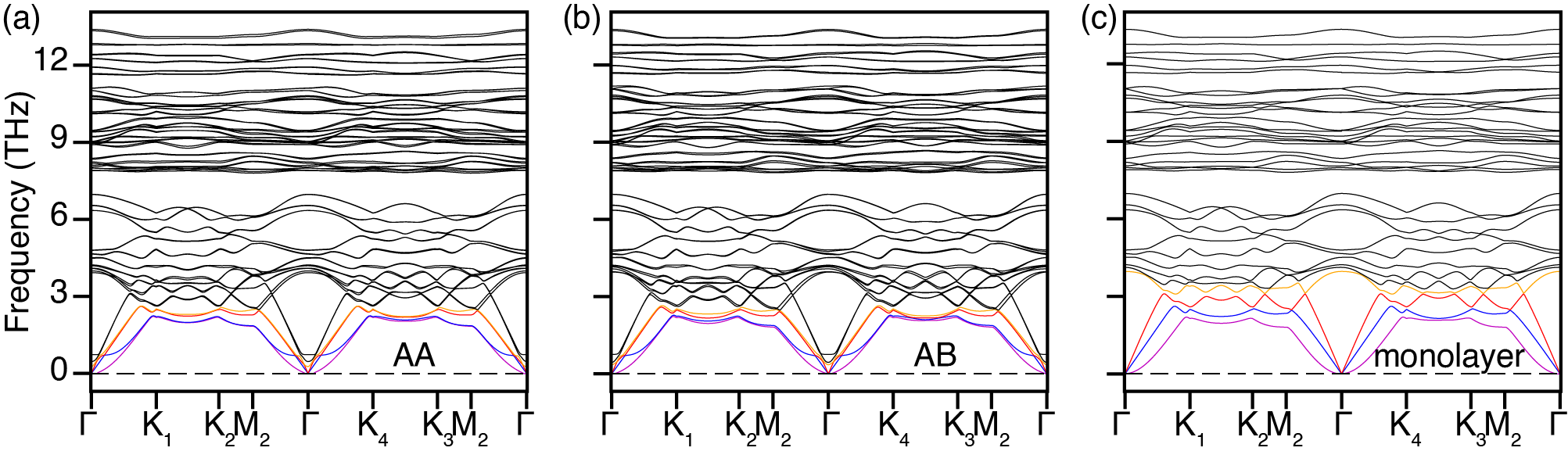}
    \caption{Phonon dispersion of ReS$_{2}$ in (a) AA stacking, (b) AB stacking, and (c) monolayer.}
    \label{structure}
\end{figure*}

\section{Low frequency Raman mode analysis}
\begin{figure*}[!htbp]
    \centering
    \includegraphics[width=0.75\linewidth]{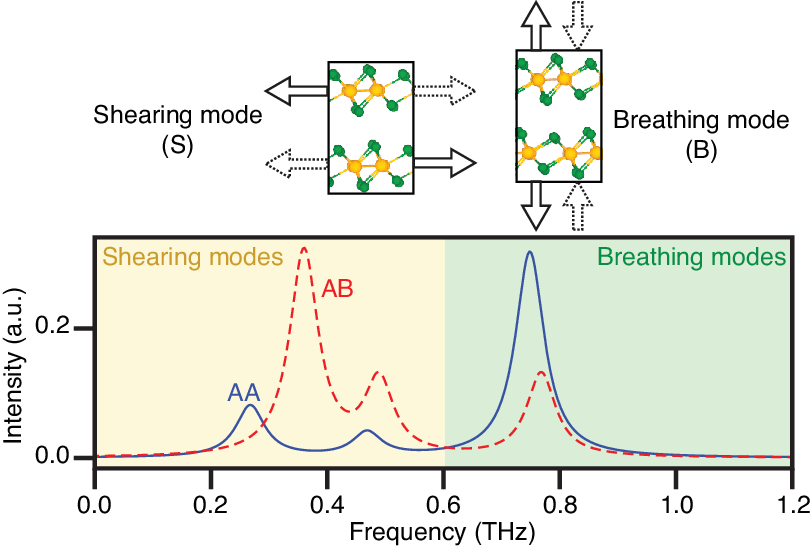}
    \caption{Raman modes in AA and AB stackings. Shear modes dominate in the lower half (yellow shaded) frequency region and the upper half by breathing modes (green shaded).}
    \label{ltc}
\end{figure*}

\end{document}